\documentclass[aps,prl,preprint,superscriptaddress]{revtex4-1}

\usepackage{graphicx}  

\begin{document}


\title{Magnetotransport in ${\bf Sr_{3}PbO}$ antiperovskite with three-dimensional massive Dirac electrons}


\author{S. Suetsugu} 
\affiliation{Department of Physics, The University of Tokyo, 7-3-1 Hongo, Bunkyo-ku, Tokyo 113-0033, Japan}
\author{K. Hayama} \affiliation{Department of Physics, The University of Tokyo, 7-3-1 Hongo, Bunkyo-ku, Tokyo 113-0033, Japan}
\author{A. W. Rost} \affiliation{Institute for Functional Matter and Quantum Technologies, University of Stuttgart, Pfaffenwaldring 57, 70550 Stuttgart, Germany} \affiliation{Max Planck Institute for Solid State Research, Heisenbergstrasse 1, 70569 Stuttgart, Germany}
\author{J. Nuss} \affiliation{Max Planck Institute for Solid State Research, Heisenbergstrasse 1, 70569 Stuttgart, Germany}
\author{C. M\"uhle} \affiliation{Max Planck Institute for Solid State Research, Heisenbergstrasse 1, 70569 Stuttgart, Germany}
\author{J. Kim} \affiliation{Max Planck Institute for Solid State Research, Heisenbergstrasse 1, 70569 Stuttgart, Germany}
\author{K. Kitagawa} \affiliation{Department of Physics, The University of Tokyo, 7-3-1 Hongo, Bunkyo-ku, Tokyo 113-0033, Japan}
\author{H. Takagi} \affiliation{Department of Physics, The University of Tokyo, 7-3-1 Hongo, Bunkyo-ku, Tokyo 113-0033, Japan} \affiliation{Institute for Functional Matter and Quantum Technologies, University of Stuttgart, Pfaffenwaldring 57, 70550 Stuttgart, Germany} \affiliation{Max Planck Institute for Solid State Research, Heisenbergstrasse 1, 70569 Stuttgart, Germany}


\date{\today}

\begin{abstract}
Novel topological phenomena are anticipated for three-dimensional (3D) Dirac electrons.
The magnetotransport properties of cubic ${\rm Sr_{3}PbO}$ antiperovskite, theoretically proposed to be a 3D massive Dirac electron system, are studied.
The measurements of Shubnikov-de Haas oscillations and Hall resistivity indicate the presence of a low density 
($\sim 1 \times 10^{18}$ ${\rm cm^{-3}}$) of holes with an extremely small cyclotron mass of 0.01-0.06$m_{e}$.
The magnetoresistance $\Delta\rho_{xx}(B)$ is linear in magnetic field $B$ with the magnitude independent of temperature.
These results are fully consistent with the presence of 3D massive Dirac electrons in ${\rm Sr_{3}PbO}$. 
The chemical flexibility of the antiperovskites and our findings in the family member, ${\rm Sr_{3}PbO}$, point to their potential as a model system in which to explore exotic topological phases.
\end{abstract}


\maketitle
Three-dimensional (3D) topological Dirac semimetals (TDS) \cite{PhysRevLett.108.140405,PhysRevB.85.195320,PhysRevB.88.125427}, 
where a 3D linear dispersion of electrons is realized in the bulk, 
have been attracting considerable interest because of their distinct expected topological properties as compared to those of two-dimensional (2D) TDS. 
3D TDS can be driven into other topological phases such as a Weyl semimetallic and a topological insulating state by breaking time reversal symmetry or by controlling the band gap \cite{PhysRevB.85.195320,PhysRevB.83.205101,yang2014classification}. 
In low carrier density systems, the quantum limit where all carriers reside in the lowest Landau level (LL) can be achieved at feasible magnetic fields. 
The quantum limit of 3D TDS is particularly attractive from the view of the physics of topological matter, as quasi one-dimensional conducting states exhibiting a chiral anomaly for example are expected \cite{nielsen1983adler,PhysRevB.88.104412}. 

${\rm Na_{3}Bi}$ \cite{PhysRevB.85.195320,liu2014discovery,xu2015observation,xiong2015evidence}, 
${\rm Cd_{3}As_{2}}$ \cite{PhysRevB.88.125427,neupane2014observation,liu2014stable,PhysRevLett.113.027603,liang2015ultrahigh,PhysRevLett.113.246402,PhysRevLett.114.117201,cao2015landau,PhysRevB.92.081306,PhysRevX.5.031037,li2015giant,li2016negative,jia2016thermoelectric}, 
${\rm Bi_{1-x}Sb_{x}}$ \cite{PhysRevLett.111.246603}, ${\rm TlBiSSe}$ \cite{PhysRevB.91.041203}, 
${\rm ZrTe_{5}}$ \cite {PhysRevB.92.075107,li2016chiral,PhysRevB.93.115414} and ${\rm Pb_{1-x}Sn_{x}Se}$ \cite{dziawa2012topological,liang2013evidence} have been experimentally confirmed to be 3D TDSs. 
The expected chiral anomaly was explored experimentally and the expected signature of a negative longitudinal magnetoresistance (MR) has been reported \cite{xiong2015evidence,li2015giant,li2016negative,li2016chiral}. 
There is however, ongoing discussion on the possible contribution of a current jetting effect on the observed negative MR \cite{arnold2016negative,dos2016search}.

\begin{figure}
\includegraphics{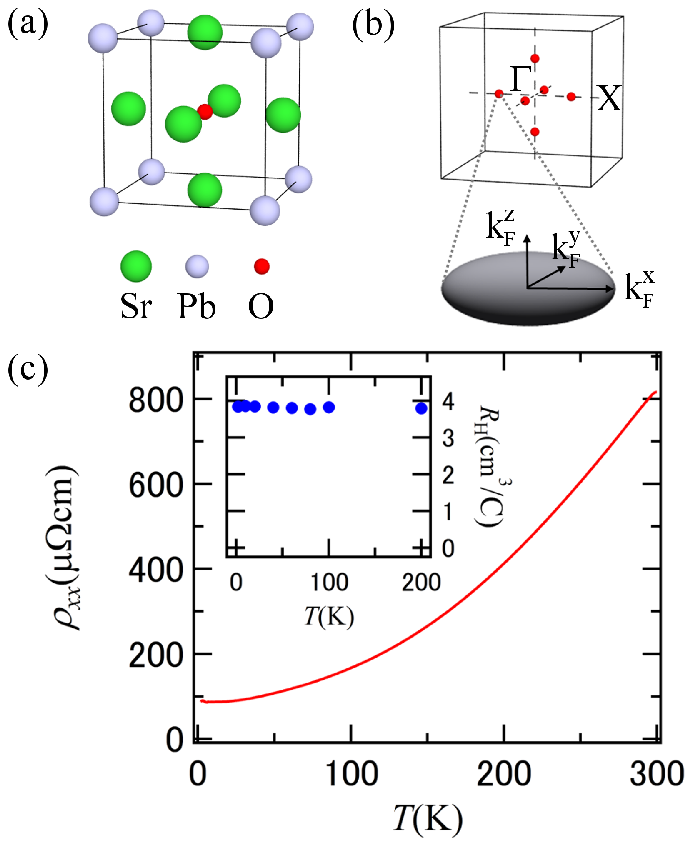}
\caption{(Color online) 
Crystal structure, Fermi surface and transport properties of ${\rm Sr_{3}PbO}$ antiperovskite. 
(a) Cubic antiperovskite structure of ${\rm Sr_{3}PbO}$. 
The positions of metal elements and O are swapped as compared to those of normal cubic perovskite oxides. 
O is at the center of ${\rm Sr_{6}}$ octahedron. 
Pb fills the space between ${\rm OSr_{6}}$ octahedra. 
(b) The first Brillouin zone of antiperovskite and the anisotropic Fermi surface on the (100) axis. 
Dirac points are located at $(\pm k_{c}, 0, 0)$, $(0, \pm k_{c}, 0)$ and $(0, 0, \pm k_{c})$ 
on the ${\rm \Gamma\mathchar`-X}$ lines which are ${\rm C_{4}}$ rotational axes in $k$ space. 
(c) Temperature dependence of resistivity $\rho_{xx}(T)$ of ${\rm Sr_{3}PbO}$ single crystal from 2 K to 300 K. 
A metallic behavior is observed. 
The inset shows the Hall coefficient $R_{H}$ in the limit of zero magnetic field, 
which is temperature-independent.
}
\label{overview}
\end{figure}

To explore those exotic phases near the 3D TDS phase further, chemically flexible 3D TDS, 
in which it is easy to break symmetries and control parameters such as the band filling, 
the spin-orbit coupling and the magnetism, are highly desirable. 
Recently, a family of antiperovskite, $A_{3}Tt{\rm O}$ ($A$ = Ca, Sr, Ba and $Tt$ = Sn and Pb), 
was theoretically proposed as a candidate system for 3D massive Dirac electrons \cite{kariyado2011three,kariyado2012low,kariyadoPhDthesis}. 
The cubic ``anti'' perovskite structure of $A_{3}Tt{\rm O}$ is shown in Fig. \ref{overview}(a). 
Here the O atom is surrounded octahedrally by $A$ atoms and forms an ${\rm O}A_{6}$ octahedron. 
The $Tt$ atom occupies the space between the ${\rm O}A_{6}$ octahedra. 

In the ionic limit, the valence state of the antiperovskites can be expressed as $A^{2+}_{3}Tt^{4-}{\rm O^{2-}}$. 
According to band calculations \cite{kariyado2011three}, 
the fully occupied $5p$ ($6p$) orbitals of $Tt^{4-}$ form valence bands and the empty $3d$ ($4d$ and $5d$) orbitals of $A^{2+}$ ions form a conduction band. 
The $d$ conduction bands and the $p$ valence bands overlap marginally, 
and the $d$-$p$ hybridization opens an energy gap at the band crossing surface. 
Only at the six equivalent points on the ${\rm \Gamma\mathchar`-X}$ lines, 
$(\pm k_{c}, 0, 0)$, $(0, \pm k_{c}, 0)$ and $(0, 0, \pm k_{c})$, 
is the crossing of the valence and conduction bands protected by the ${\rm C_{4}}$ rotational symmetry, 
which leads to six Dirac points around which anisotropic 3D Dirac bands are found (Fig. \ref{overview}(b)). 
Recent angle-resolved photoemission spectroscopy study on ${\rm Ca_{3}PbO}$ was consistent with the band calculation though hole doping limited the observation of bands only well below the Dirac points \cite{obata2017arpes}.

The 3D Dirac band in the antiperovskites family is shown to have a small mass gap of the order of 10 meV \cite{kariyado2011three}. 
The mass gap originates from the contribution of high energy orbital states and the spin-orbit coupling. 
The magnitude of the gap therefore depends on $A$ and $Tt$ elements. 
The presence of a tunable mass gap may give rise to an even richer variety of topological states. 
${\rm Ca_{3}PbO}$ and ${\rm Sr_{3}PbO}$ were predicted to be a topological crystalline insulator (TCI) \cite{hsieh2014topological}, 
where type-II Dirac surface states were recently predicted \cite{PhysRevB.95.035151}. 
The cubic perovskite structure is stable over a wide range of $A$ and $Tt$ atoms; 
$A$ = Ca, Sr, Ba, and magnetic Eu and $Tt$ = Pb and Sn \cite{nuss2015tilting}. 
The chemical flexibility should make the antiperovskites a unique and promising material family for band engineering of 3D Dirac electrons. 

Here, we report the magnetotransport studies of a single crystal of ${\rm Sr_{3}PbO}$, one of the cubic antiperovskites. 
Our analysis of Shubnikov-de Haas (SdH) oscillations indicates an extremely small effective mass of the lightly naturally doped holes, $n \sim 10^{18}$ ${\rm cm^{-3}}$. 
Further, a transverse magnetoresistance (MR) almost linear in magnetic field ($B$) and temperature ($T$) independent was observed. 
These results are entirely consistent with the existence of 3D massive Dirac electrons in ${\rm Sr_{3}PbO}$ antiperovskite. 

Single crystals of ${\rm Sr_{3}PbO}$ were grown by a grain growth enhanced with 3-5\% excess of alkaline earth elements in a sealed Ta ampoule as reported previously \cite{nuss2015tilting}. 
The detailed structural characterization of single crystal prepared in the same manner is described in Ref. \cite{nuss2015tilting}. 
Because ${\rm Sr_{3}PbO}$ crystals are extremely air sensitive, 
all the preparations for the transport measurements were conducted inside an Ar-filled glove box. 
The crystals were polished into a rectangular shape and placed in a sample holder with a vacuum tight indium seal. 
To ensure good electrical contacts, a gold film was coated to serve as an electrode, 
where a gold wire was attached using conductive epoxy cw2400 (Circuit Works) with the help of a micromanipulator.

The magnetotransport measurements were performed using PPMS 14 T (Quantum Design) with magnetic fields perpendicular to the applied current. 
A six-probe configuration was used to measure the resistivity $\rho_{xx}(B)$ and the Hall resistivity $\rho_{xy}(B)$. 
For all measurements reported here, we confirmed that the results for the two independent pairs were consistent with each other and that any artifact originating from spatial inhomogeneity of the sample is eliminated. 
To exclude the contribution of $\rho_{xy}(B)$ ($\rho_{xx}(B)$) component to $\rho_{xx}(B)$ ($\rho_{xy}(B)$) due to the misalignment of electrodes, 
$B$-symmetric ($B$-asymmetric) components of the raw data were numerically calculated and presented as $\rho_{xx}(B)$ ($\rho_{xy}(B)$) in this paper. 
Because of the air sensitivity, we could not take the crystals out from the glove box for singe crystal x-ray measurements. 
Observation of $\sim$ 5 T frequency of quantum oscillations, 
in conjunction with the result of the magnetic torque measurements \cite{JKimthesis}, 
indicates that the direction of magnetic field in the present experiments is close to (100) axis.

The measurements were conducted over five single crystals from two different batches, including the one of which we showed the data in the main text. 
The result was reasonably reproducible, as shown in the Supplemental Material \cite{supplement}. 
Bulk superconductivity, as reported in polycrystalline ${\rm Sr_{3-x}SnO}$ \cite{oudah2016superconductivity}, was not observed in the single crystals used in this study down to 2 K. 
A trace of filamentary superconductivity at low temperatures was often observed (See Supplemental Material for details \cite{supplement}).

\begin{figure}
\includegraphics{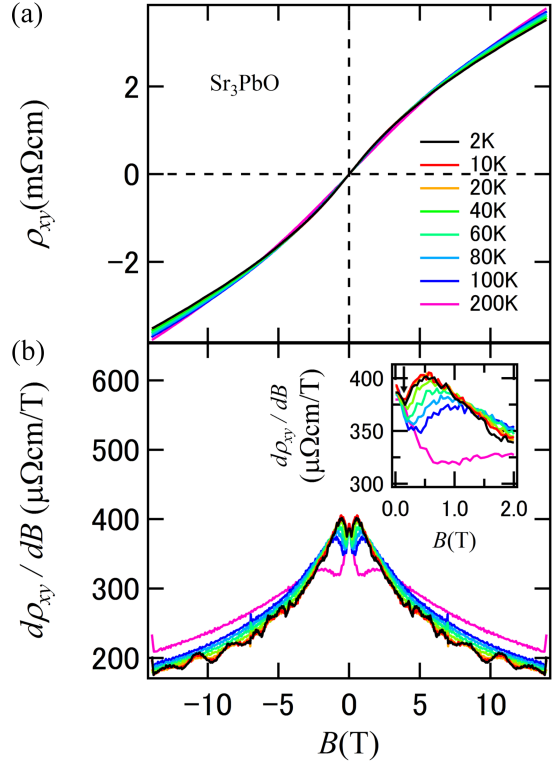}
\caption{(Color online) 
Hall resistivity $\rho_{xy}(B)$ of antiperovskite ${\rm Sr_{3}PbO}$ single crystal. 
(a) Magnetic field ($B$) dependence of Hall resistivity $\rho_{xy}(B)$, is non-linear in $B$ and not strongly $T$-dependent at least up to 200 K. 
Positive and $T$-independent $R_{H} = +3.8$ ${\rm cm^{3}/C}$ is observed in the zero field limit, 
yielding a hole density of $1.6 \times 10^{18}$ ${\rm cm^{-3}}$. 
(b) The non-linear behavior of $\rho_{xy}(B)$ is better recognized in the derivative of Hall resistivity $d\rho_{xy}/dB$. 
A crossover from low field $T$-independent to high field $T$-independent behavior can be seen in the inset.
}
\label{Hall}
\end{figure}

$T$-dependent resistivity $\rho_{xx}(T)$ for the ${\rm Sr_{3}PbO}$ single crystal shows metallic behavior up to room temperature 
with a residual resistivity $\rho_{xx}(0) = 90$ ${\rm \mu\Omega cm}$ and a residual resistance ratio RRR of $\sim$ 10 as seen in Fig. \ref{overview}(c). 
The Hall resistivity $\rho_{xy}(B)$ (Fig. \ref{Hall}(a)) and the derivative of Hall resistivity $d\rho_{xy}/dB$ (Fig. \ref{Hall}(b)) 
in the zero-field limit give a positive and $T$-independent $R_{H} = + 3.8$ ${\rm cm^{3}/C}$ (inset in Fig. \ref{overview}(c)), 
yielding a hole density of $1.6 \times 10^{18}$ ${\rm cm^{-3}}$. 
$\rho_{xy}(B)$ in Fig. \ref{Hall}(a) is non-linear in $B$, which is reminiscent of systems with two types of carriers (See also Supplemental Material \cite{supplement}). 
The mass anisotropy of six hole-pockets, which we will discuss later, 
may give rise to the coexistence of high and low mobility holes. 
Within the two-carrier model, the temperature independence of the low field limit $R_{H}$ suggests that 
only the high mobility holes dominate $\rho_{xy}(B)$ in the zero-field limit. 
The Hall mobility is estimated to be $4.4 \times 10^{4}$ ${\rm cm^{2}/Vs}$ in the low $T$ limit, 
comparable to those reported for other 3D Dirac systems, 
${\rm Cd_{3}As_{2}}$ \cite{PhysRevLett.113.246402,PhysRevLett.114.117201} and ${\rm ZrTe_{5}}$ \cite{PhysRevB.93.115414}. 
The high mobility of carriers is indicative of the presence of 3D Dirac electrons. 
The hole-doping is naturally expected from the viewpoint of chemistry. 
The preferred valence states of Pb are 2+ and 4+.
The extremely reduced ionic state of ${\rm Pb^{4-}}$ should favor an oxidation of the sample, 
for example through cation defects or excess oxygens, which should result in hole-doping.

\begin{figure}
\includegraphics{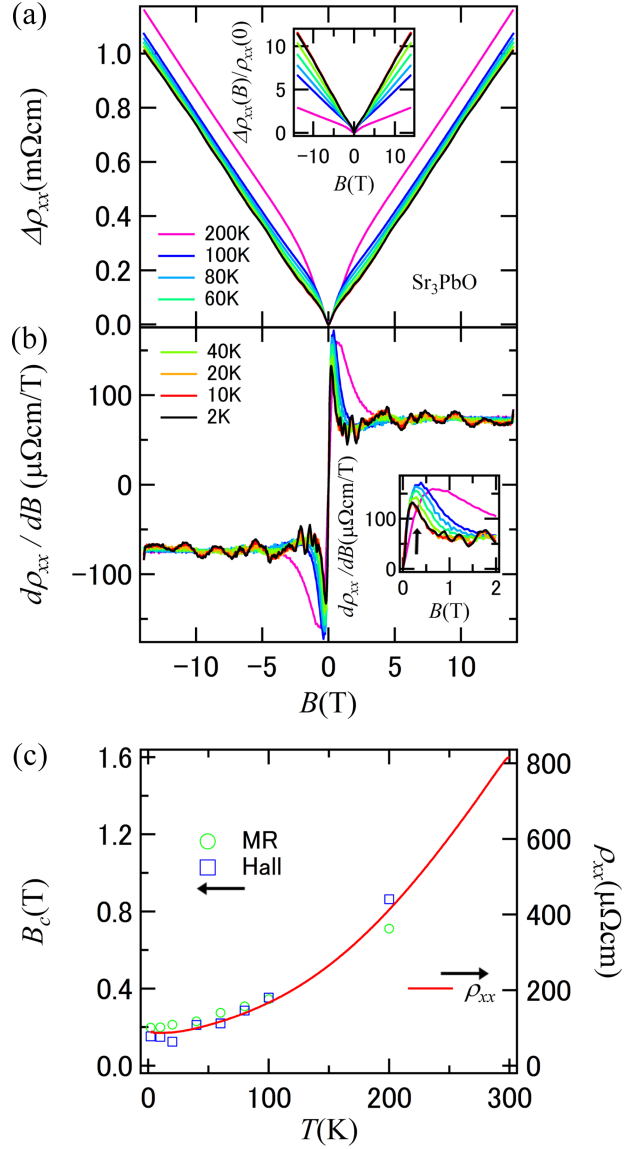}
\caption{(Color online) 
Magnetoresistance and its crossover of ${\rm Sr_{3}PbO}$. 
(a) Magnetic field ($B$) dependence of $\Delta\rho_{xx}(B) = \rho_{xx}(B) - \rho_{xx}(0)$ 
and magnetoresistance ratio $\Delta\rho_{xx}(B)/\rho_{xx}(0)$. 
At $T = 2$ K, a large magnetoresistance ratio over 10 at $B = 14$ T is observed as seen in the inset. 
A $B$-linear behavior is observed at high fields. 
The $B$-linear contribution of $\Delta\rho_{xx}(B)$ is $T$-independent up to 200 K at high fields. 
(b) $T$-independent and $B$-linear behavior of $\Delta\rho_{xx}(B)$ is better recognized 
in $T$-independent and constant derivative of resistivity $d\rho_{xx}/dB$ at high field. 
SdH oscillations are clearly observed above 1 T. 
A crossover from $B^{2}$ behavior to $B$-linear behavior in $\Delta\rho_{xx}(B)$ is emphasized in the inset. 
(c) The crossover fields $B_{c}$ defined by the peak in $d\rho_{xx}/dB$ denoted by the black arrow in Fig. \ref{MR}(b) (circle) 
and by the dip in $d\rho_{xy}/dB$ denoted by the black arrow in Fig. \ref{Hall}(b) (square) 
seem to be scaled with $\rho_{xx}(T)$ (line).
}
\label{MR}
\end{figure}

At the lowest temperature measured, $T = 2$ K, a large MR ratio $\Delta\rho_{xx}(B)/\rho_{xx}(0)$ of over 10 at $B = 14$ T is observed as seen in the inset to Fig. \ref{MR}(a) (See also Supplemental Material \cite{supplement}). 
Over a wide range of magnetic field $B$ from 1 T up to 14 T, MR shows almost $B$-linear behavior. 
No trace of saturation is seen up to 14 T. 
This behavior has been commonly observed in other 3D TDS \cite{PhysRevLett.113.246402,PhysRevLett.114.117201,liang2015ultrahigh,PhysRevB.92.081306,PhysRevX.5.031037,PhysRevB.91.041203,xiong2015evidence} 
and topological semimetals \cite{PhysRevB.84.220504,wang2013large,pavlosiuk2015shubnikov}. 
The close similarity to other 3D TDS again supports the presence of 3D Dirac electrons in ${\rm Sr_{3}PbO}$.

While $\Delta\rho_{xx}(B)/\rho_{xx}(0)$ shows a decrease with increasing $T$, $\Delta\rho_{xx}(B)$, 
more specifically the $B$-linear contribution of $\Delta\rho_{xx}$ at high fields, 
is surprisingly independent of $T$ up to 200 K as seen in Fig. \ref{MR}(a). 
This can be better visualized as the $T$- and $B$- independent derivative of resistivity $d\rho_{xx}/dB$ in Fig. \ref{MR}(b) (See also Supplemental Material \cite{supplement}).
$\Delta\rho_{xx}(B)/\rho_{xx}(0)$ in the $B$-linear region is therefore scaled by $B/\rho_{xx}(0,T)$ and a Kohler's rule, 
$\Delta\rho_{xx}(B,T)/\rho_{xx}(0,T) = f(B/\rho_{xx}(0,T))$, holds for the high field $B$-linear contribution (See Supplemental Material for details \cite{supplement}). 

In the low field limit, $\Delta\rho_{xx}(B)$ shows $B^{2}$-behavior demonstrated by almost $B$-linear behavior of $d\rho_{xx}/dB$. 
Interestingly, the slopes of $d\rho_{xx}/dB$ for different temperatures overlay each other as seen in the inset to Fig. \ref{MR}(b). 
This means that the magnitude of the $B^{2}$-contribution of $\Delta\rho_{xx}(B)$ in the zero-field limit 
is also almost $T$-independent. 
In contrast to the high field $B$-linear contribution, 
the low field $B^{2}$-contribution apparently violates Kohler's rule and cannot be captured simply by a classical $B^{2}$ MR. (See Supplemental Material for details \cite{supplement})

A crossover from the low field $B^{2}$ to the high field $B$-linear behaviors in $\Delta\rho_{xx}(B)$ can be seen in $d\rho_{xx}/dB$ in the inset of Fig. \ref{MR}(b). 
A crossover field $B_{c}^{\rm MR}$ can be represented by the magnetic field where $d\rho_{xx}/dB$ shows a peak (black arrow in the inset to Fig. \ref{MR}(b)). 
$B_{c}^{\rm MR}$ increases systematically with $T$, as seen in Fig. \ref{MR}(c). 

The non-linear magnetic field dependence of Hall resistivity $\rho_{xy}(B)$ in Fig. \ref{Hall}(a) clearly mirrors the crossover from $B^{2}$ to $B$ behavior observed in $\Delta\rho_{xx}(B)$. 
The non-linearity is more clearly distinguished in the derivative $d\rho_{xy}/dB$ shown in Fig. \ref{Hall}(b). 
In $d\rho_{xy}/dB$, a crossover is observed from a $B$-linear and $T$-independent decrease well below 1 T 
to a weak and again $T$-independent decrease at high fields, as seen in the inset to Fig. \ref{Hall}(b). 
The crossover magnetic field $B_{c}^{\rm Hall}$ may be represented by the position of the dip in $d\rho_{xy}/dB$ (black arrow in the inset to Fig. \ref{Hall}(b)), 
which well agrees with the corresponding crossover field $B_{c}^{\rm MR}$ for $\Delta\rho_{xx}(B)$ as shown in Fig. \ref{MR}(c). 
The crossover $B_{c}$ between the two $T$-independent magnetotransport regions 
commonly observed in $\Delta\rho_{xx}(B)$ and $\rho_{xy}(B)$ is scaled with $\rho_{xx}(T)$ 
and hence the $T$-dependent scattering rate $1/\tau$, which means that the crossover is controlled by $B\tau (T)$.

\begin{figure}
\includegraphics{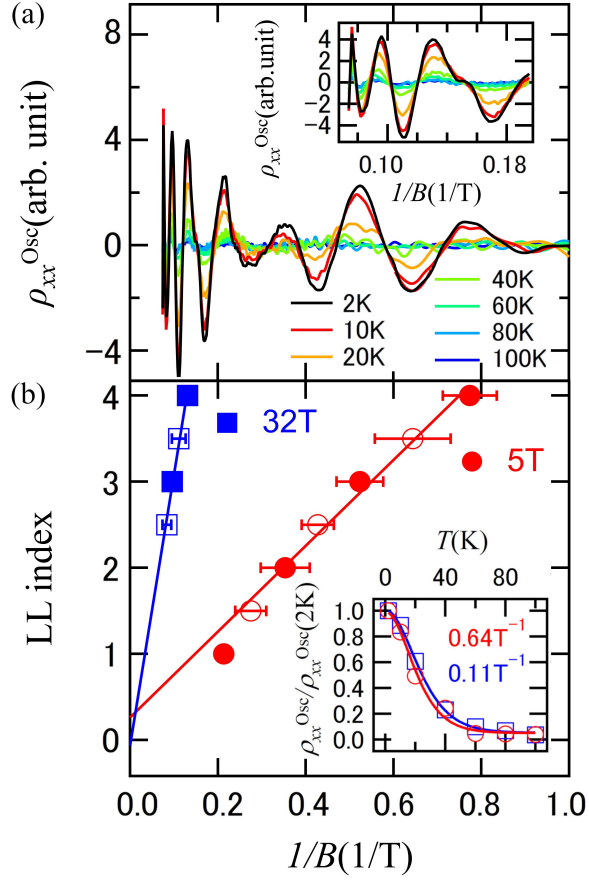}
\caption{(Color online) 
SdH oscillations of ${\rm Sr_{3}PbO}$. 
(a) SdH oscillations are derived from $\rho_{xx}^{\rm Osc}(B)$ obtained 
by numerically subtracting a polynomial background from $\rho_{xx}(B)$. 
Two sets of oscillations are observed with frequencies 5 T and 32 T. 
The oscillations with a frequency of 32 T above 6 T are emphasized in the inset. 
(b) Landau fan diagram of oscillations with frequencies of 5 T and 32 T at 2 K to extract the phase offsets. 
Both peaks (closed symbol) and dips (open symbol) of SdH oscillations are plotted. 
Extrema of SdH oscillations and their errors were determined by gaussian fitting of SdH oscillation peaks. 
The linear fit to the 5 T oscillations for $n \geq 1.5$ and to the 32 T oscillations for $n \geq 2.5$ 
yields $\beta = 0.76 \pm 0.16$ and $\beta = 0.44 \pm 0.26$ respectively. 
The inset shows Lifshitz-Kosevitch fitting to derive the cyclotron effective mass $m_{c}^{*}$. 
The oscillation amplitudes at 0.64 ${\rm T^{-1}}$ (black arrow in (a)) and 0.11 ${\rm T^{-1}}$ (black arrow in the inset to (a)) are 
plotted as a function of temperature, representing 5 T and 32 T oscillation. 
The extracted $m_{c}^{*}$ for 0.64 ${\rm T^{-1}}$ and 0.11 ${\rm T^{-1}}$ are extremely light values of $0.011m_{e}$ and $0.057m_{e}$ respectively.
}
\label{SdH}
\end{figure}

In $\rho_{xx}(B)$, SdH oscillations are superposed at least up to 40 K. 
The observation of the SdH signal up to a high $T$ implies an extremely light carrier mass, consistent with transport due to Dirac electrons. 
Fig. \ref{SdH}(a) shows the SdH oscillations after subtracting a polynomial background from $\rho_{xx}(B)$ (See also Supplemental Material \cite{supplement}). 
The oscillations are composed of two different traces with frequencies of $4.98 \pm 0.30$ T and $31.5 \pm 2.4$ T \cite{comment} hereinafter referred as 5 T and 32 T respectively. 
For 5 T oscillations, the quantum limit is reached above $B$ $\sim$ 5 T, 
which is lower than or at least comparable to those reported for known 3D TDS \cite{xiong2015evidence,PhysRevX.5.031037,PhysRevB.93.115414,liang2013evidence}. 
This means that the antiperovskite ${\rm Sr_{3}PbO}$ is a promising arena for physics in the quantum limit. 
Oscillations with frequency 32 T persist up to higher fields than 14 T and the quantum limit cannot be reached. 
Angle dependent magnetic torque oscillation measurements \cite{JKimthesis} indicate that 
the magnetic field orientation of our sample is along a high symmetry ([100]) direction. 

The cyclotron effective masses $m_{c}^{*}$ for the two oscillations are derived by 
fitting the temperature dependence of the magnitude of SdH oscillations to the Lifshitz-Kosevich equation \cite{lifshitz1956theory} as shown in the inset to Fig. \ref{SdH}(b). 
The clear oscillation peaks at 0.64 ${\rm T^{-1}}$ (black arrow in Fig. \ref{SdH}(a)) and 0.11 ${\rm T^{-1}}$ (black arrow in the inset to Fig. \ref{SdH}(a)) are used for the fitting. 
This yields a $m_{c}^{*}$ of $0.011m_{e}$ and $0.057m_{e}$ for the 5 T and 32 T oscillations respectively. 
The extremely light effective mass of as low as one percent of the free electron mass is consistent with the presence of the Dirac electrons. 
The difference in $m_{c}^{*}$ between the two oscillation frequencies may be attributed to the anisotropy of cyclotron orbits within the hole pockets around the Dirac points, which will be discussed later.

The nontrivial Berry phase of Dirac electrons may be reflected in an extra phase offset of SdH oscillations. 
In an ideal Dirac electron system the additional offset phase $\beta$ is related to the $\pi$ Berry phase associated with the cyclotron motion \cite{novoselov2005two,zhang2005experimental,PhysRevLett.113.246402}. 
Indeed, $\beta = 0.5$ is expected for an ideal Dirac dispersion and $\beta = 0$ for a trivial $k$-parabolic band \cite{PhysRevLett.82.2147}. 
This offset can be extracted from the Landau fan diagram by a linear fit of the position of maxima 
in SdH oscillations in $\sigma_{xx}$ with $n = F/B - 1/2 + \beta - 1/8$, 
where $n$ and $F$ are the maximum index and the frequency of SdH oscillations, respectively. 
The additional factor $-1/8$ reflects the three-dimensionality and the maximal cross-sectional area \cite{lifshitz1956on,PhysRevLett.93.166402,PhysRevLett.97.256801}. 
As a consequence of the fact that $|{\sigma}_{xy}| > {\sigma}_{xx}$ 
and hence ${\rho}_{xx} \simeq {\sigma}_{xx}/{\sigma}_{xy}^{2}$ holds in the field range of the SdH oscillations, 
it is appropriate to assign the maxima of $\sigma_{xx}$ with the maxima of $\rho_{xx}$ (See also Supplemental Material \cite{supplement}) \cite{PhysRevB.86.045314}.

Such fits to the Landau fan diagram for the two frequencies observed by us are presented in Fig. \ref{SdH}(b). 
In order to avoid deviations from linearity arising in the quantum limit \cite{PhysRevB.84.035301} these were limited to $n \geq 1.5$. 
The resulting values of $\beta = 0.76 \pm 0.16$ and $\beta = 0.44 \pm 0.26$ for the 5 T and 32 T oscillations respectively are incompatible with the trivial value of 0, 
although they deviate from 0.5 expected for ideal Dirac dispersions. 
Such deviations have also been seen in e.g., ${\rm Cd_{3}As_{2}}$ \cite{cao2015landau} 
and can be attributed to the relevance of quadratic terms not considered in the ideal Dirac equation 
as well as the significance of spin splitting at high magnetic fields \cite{PhysRevB.84.035301,alexandradinata2017modern,PhysRevB.87.085411}. 
However, large errors involving the determination of SdH oscillation peaks make it difficult to extract clear conclusion about the offset in Landau levels.
A more detailed analysis of the phase has therefore to be referred to a future study.

Let us discuss the Fermi surface (FS) geometry and the relevant physical parameters of the antiperovskite from SdH oscillations. 
We assume that the FS has an ellipsoidal shape and the magnetic field has been applied close to the (100) axis. 
The three principal Fermi momenta $k_{F}^{i}$ ($i = x, y, z$) are defined for the FS on the (100) axis as in Fig. \ref{overview}(b). 
The (100) axis has ${\rm C_{4}}$ rotational symmetry and thus $k_{F}^{y} = k_{F}^{z}$. 
There should be two cross sectional areas, $\pi k_{F}^{y}k_{F}^{z}$ for the two FSs on the (100) axis 
and $\pi k_{F}^{x}k_{F}^{y} = \pi k_{F}^{x}k_{F}^{z}$ for the other four FSs on (010) and (001). 
The SdH frequencies, 5 T and 32 T, yield the cross sectional area of the FS normal to $B$ as $5.0 \times 10^{-4}$ \AA$^{-2}$ and $2.8 \times 10^{-3}$ \AA$^{-2}$, respectively. 
If we assign the small (5 T) and the large (32 T) FS cross sections to the areas $\pi k_{F}^{y}k_{F}^{z}$ and $\pi k_{F}^{x}k_{F}^{y}$ respectively, 
we obtain $k_{F}^{x} = 5.6k_{F}^{y} = 0.073$ \AA$^{-1}$. 
These $k_{F}$ values yield a carrier density $n_{\rm SdH} = 6 \times k_{F}^{x}k_{F}^{y}k_{F}^{z} / 3\pi^{2} = 2.5 \times 10^{18}$ ${\rm cm^{-3}}$, 
which is close to the Hall carrier density $n_{\rm Hall} = 1.6 \times 10^{18}$ ${\rm cm^{-3}}$. 
Note that if we assume a compressed FS instead of one that is elongated along the (100) axis, 
$5.6k_{F}^{x} = k_{F}^{y} = 0.030$ \AA$^{-1}$ and $n_{\rm SdH} = 1.0 \times 10^{18}$ ${\rm cm^{-3}}$ result. 
We cannot rule out at this stage the possibility of compressed FS as the agreement 
between $n_{\rm SdH}$ and $n_{\rm Hall}$ for compressed FS is not entirely unreasonable given the uncertainty of the relevant parameters.

Using $E_{F} = m_{c}^{*}v_{F}^{y}v_{F}^{z} = m_{c}^{*} E_{F}/\hbar k_{F}^{y} \times E_{F}/\hbar k_{F}^{z}$, 
we estimate a Fermi energy of $E_{F} = 117$ meV 
and Fermi velocities of $5.6v_{F}^{x} = v_{F}^{y} = v_{F}^{z} = 1.4 \times 10^{6}$ m/s from the 5 T oscillations with cyclotron effective mass $0.011m_{e}$. 
The same analysis applies for 32 T oscillations and gives consistent numbers, $E_{F} = 127$ meV 
and $5.6v_{F}^{x} = v_{F}^{y} = v_{F}^{z} = 1.5 \times 10^{6}$ m/s. 
The energy of the $n = 1$ LL mode $E_{1}$, for 5 T oscillations is estimated to be 114 meV 
using $E_{n} = \sqrt{2v_{F}^{y}v_{F}^{z}\hbar eBn}$. 
$E_{1} \sim E_{F}$ at 5 T is consistent with the realization of the quantum limit at $\sim$ 5 T. 
If we choose the case for compressed FS shape, the same analysis yields $E_{F} = 111$ meV, 
$v_{F}^{x}/5.6 = v_{F}^{y} = v_{F}^{z} = 0.6 \times 10^{6}$ m/s and $E_{1} = 115$ meV, 
which is also consistent with the quantum limit at $\sim$ 5 T. 
Note that a mass gap of $\Delta \sim 10$ meV obtained from a band calculation \cite{kariyado2011three,kariyadoPhDthesis} 
can be ignored in the estimates of $E_{F}$ and $E_{1}$, as $\Delta$ is much smaller than the calculated $E_{F}$ and $E_{1}$. 
The fact that the two cross sectional areas observed in SdH oscillations give almost the same $E_{F}$ value supports 
the assumption that they originate from the equivalent FSs. 
The angular dependence of the de Haas van Alphen signal around (100) observed by torque measurement on single crystals with a smaller carrier concentration than the present crystals gives an ``extrapolated'' anisotropy of 2-3 \cite{JKimthesis}, 
a factor of two smaller than the present estimate. 
A band calculation gives an estimate of anisotropy of 2 \cite{kariyadoPhDthesis}. 
The difference may suggest the presence of more complicated FS shape than ellipsoid for a higher carrier concentration, 
which should be clarified in future.

To understand the $B$-linear MR characteristic to Dirac systems, 
two models are often employed but, at least in their original form, do not provide full account for the experimental observations in this study. 
One is the classical disorder model discussed in doped silver chalcogenides \cite{parish2003non,PhysRevB.75.214203}, ${\rm Cd_{3}As_{2}}$ \cite{PhysRevLett.114.117201} and GaAs quantum well \cite{PhysRevLett.117.256601}, 
where $B$-linear $\Delta\rho_{xx}(B)$ originates from $B$-linear $\rho_{xy}(B)$. 
The observed close link between $\Delta\rho_{xx}(B)$ and $\rho_{xy}(B)$ via mobility supports this scenario.
However, it is not obvious at all in this scenario why there is a $T$-independence of $\Delta\rho_{xx}(B)$ and $\rho_{xy}(B)$ 
in conjunction with a $T$-dependent $\rho_{xx}(B)$, 
and the absence of non-linearity in $\Delta\rho_{xx}(B)$ out of the non-linear $\rho_{xy}(B)$.
Temperature independence of $\Delta\rho_{xx}(B)$ might suggest the static impurity scattering. 
The other scenario is based on the quantum limit behavior of Dirac electrons \cite{PhysRevB.58.2788}. 
This may account for the $T$-independent and $B$-linear $\Delta\rho_{xx}(B)$. 
The crossover to $B$-linear behavior, however, seems to be controlled by $B\tau$ rather than the quantum limit of $B = 5$ T. 
With the application of magnetic field along (100), 
the six equivalent FSs may split into the two groups due to the FS anisotropy. 
The two contributions can have different mobilities under magnetic fields 
but the very weak temperature dependence of $\Delta\rho_{xx}(B)$ and $\rho_{xy}(B)$ makes the analysis 
in terms of the na\"ive two carrier model and hence the interpretation of $\Delta\rho_{xx}(B)$ and $\rho_{xy}(B)$ very difficult. 
We at this stage cannot exclude the possibility that the FS anisotropy can account for the difficulties of the two scenarios. 
Measurements with magnetic fields parallel to the (111) direction, where all the six FSs are equivalent, 
may help in discriminating between the two scenarios by eliminating the complications arising from FS anisotropy.

In summary, we report the magnetotransport properties of ${\rm Sr_{3}PbO}$, 
a putative 3D massive Dirac electron system and one of the members of the antiperovskite family. 
The analysis of SdH oscillations reveals an extremely small cyclotron mass of hole carriers as small as 1\% of the free electron mass, 
which supports the presence of 3D massive Dirac electrons in ${\rm Sr_{3}PbO}$. 
Exotic properties of Dirac electrons may be reflected as $T$-independent and $B$-linear $\Delta\rho_{xx}(B)$. 
The quantum limit is achieved in some of the Fermi pockets at fields as low as $B = 5$ T. 
Our results open up a new material family to design topologically distinct phases derived from the 3D Dirac electrons and to explore exotic phases anticipated in the quantum limit.

\begin{acknowledgments}
We thank T. Kariyado, A. Schnyder, A. Yaresko, M. G. Yamada and M. Ogata for discussion, and K. Pflaum and M. Dueller for technical assistance.
This work was partly supported by Japan Society for the Promotion of Science (JSPS) KAKENHI (No. 24224010, 15K13523, JP15H05852, JP15K21717, 17H01140) and Alexander von Humboldt foundation.
\end{acknowledgments}

\end{document}